# Soft x-ray irradiation induced metallization of layered TiNCl

Noriyuki Kataoka[1], Masashi Tanaka[2], Wataru Hosoda[1], Takumi Taniguchi[1], Shin-ichi Fujimori[3], Takanori Wakita[1,4], Yuji Muraoka[1,4], and Takayoshi Yokoya[1,4]

1. Graduate School of Natural Science and Technology, Okayama University, Okayama 700-8530, Japan
2. Graduate School of Engineering, Kyushu Institute of Technology, Kitakyushu 804-8550, Japan
3. Materials Sciences Research Center, Japan Atomic Energy Agency, Sayo, Hyogo 679-5148, Japan
4. Research Institute for Interdisciplinary Science, Okayama University, Okayama 700-8530, Japan

Abstract

   We have performed soft x-ray spectroscopy in order to study the photoirradiation time dependence of the valence band structure and chemical states of layered transition metal nitride chloride TiNCl. Under the soft x-ray irradiation, the intensities of the states near the Fermi level ($E_F$) and the $Ti^{3+}$ component increased, while the Cl $2p$ intensity decreased. Ti $2p$-$3d$ resonance photoemission spectroscopy confirmed a distinctive Fermi edge with Ti $3d$ character. These results indicate the photo-induced metallization originates from deintercalation due to Cl desorption, and thus provide a new carrier doping method that controls the conducting properties of TiNCl.

**Introduction**

   TiNCl is an interesting layered material that belongs to a family of layered nitride halides $MNX$ ($M$ = Ti, Zr, Hf and $X$ = Cl, Br, I) systems with two polymorphs, the FeOCl-type crystal structure (α-form) and the SmSI-type crystal structure (β-form) [1, 2]. TiNCl is known to take only the FeOCl-type crystal structure [2], consisting of a stack of orthogonal $M$-N layers located between two $X$ layers. TiNCl is a band insulator with a direct band gap of approximately 0.5-3 eV [3, 4], and theoretical studies have proposed its application to optoelectronic devices [4], photocatalysts [5], and spin devices [6]. Upon Na intercalation between TiNCl layers, TiNCl becomes a superconductor with a relatively high superconducting transition temperature of ~16 K [7]. As electron-doped β-form HfNCl and ZrNCl [8-20], electron-doped TiNCl is considered a candidate for unconventional superconductors, where exotic mediation forces for Cooper pairing other than phonon are discussed [21-25].

In order to further explore the physical properties of this remarkable material, it is essential to control carrier concentrations. In the β-form *M*N*X*, the carrier control has been performed by intercalation of an alkali- or alkaline-earth-metal [8, 9, 15, 16], and/or off stoichiometry or deintercalation of *X* [26]. As for TiNCl, to the best of our knowledge, carrier control by neither element substitution nor off stoichiometry has been reported. This may be due to the fact that TiNCl is easily thermally decomposed into TiN by annealing[27-29]. The intercalation of alkali metal atoms and/or organic molecules between layers is the only method way for doping carriers. However, intercalated samples are unstable in a humid air, and precise carrier controlling is challenging, which prevents the systematic investigations of physical properties systematically.

One of the techniques to create conductive samples is photo-irradiation-induce metallization by breaking chemical bonds [30-36]. In $SrTiO_3$ and $TiO_2$, a metallic state called two-dimensional electron gases is created on the surface by photoirradiation. Two-dimensional electron gases have attracted extensive attention owing to unique physical properties [30-35, 37-39], providing opportunities for the development of next generation of electronic and photonic devices. If photo-induced bond breaking is also effective in TiNCl, it can be established as a new method to control the physical properties of TiNCl. Therefore, it is one of the important challenges to investigate whether TiNCl is metallized by photoirradiation.

In this paper, we performed valence and core level photoelectron spectroscopy (PES), x-ray absorption spectroscopy (XAS), and Ti $2p$-$3d$ resonance photoelectron spectroscopy (RPES) of layered transition metal nitride chloride TiNCl. It was found that soft x-ray irradiation induced the metallization of TiNCl, and the metallization was closely related to the Cl desorption. This is the first report of electron-doped TiNCl due to Cl desorption.

**Experiment**

The highly crystalline TiNCl was grown by a method described elsewhere [7]. It was pelletized into disk shape in an Ar filled glove box. Subsequently, the pellet was transferred to an ultra-high vacuum chamber for photoelectron spectroscopy without being exposed to the humid air for a long time. The experiments were performed at the soft x-ray beam line BL23SU of SPring-8 [40] using a photoemission spectrometer equipped with a Gammadata-Scienta SES-2002 electron analyzer. In the RPES measurement, the energy resolution was set to 80 meV, and in the other PES measurement, it was set to less than 200 meV. To reduce a possibility for pyrolysis of TiNCl, we cooled down the sample and measured it at 100 K instead of at 300 K. We performed PES measurements using a photon energy of 1000 eV. For calibration of the binding energy, we used the Fermi edge of a gold film, which was located close to the sample and had a good electrical contact to the sample.

The irradiation light of $h\nu = 1000$ eV was used, and the detector angle was set to 90°. Sets of Ti $2p$, N $1s$, Cl $2p$, and valence spectra were also measured at $h\nu = 1000$ eV in the time interval of 25 minutes without changing the measured location of the sample surface. The irradiation time includes the time of the PES measurements. Though it took 8 min to take a set of Ti, N, Cl, and valence spectra, we refer to the irradiation time as the time when we started the measurement of Ti. The spot size of the excitation light was $100 \times 200$ $\mu m^2$, and the photon flux was $2 \times 10^{12}$ Photons/sec [40]. The illustration of the experimental geometry is shown in Fig. 1. The intensity of PES spectra was normalized by the area intensity of Ti $2p$ core-level spectra, while those of RPES were normalized by the photocurrent. XAS spectrum was measured using the total electron yield (TEY) mode. To obtain a clean surface, the TiNCl pellet was fractured at 100 K under an ultrahigh vacuum of $1.5 \times 10^{-8}$ Pa.

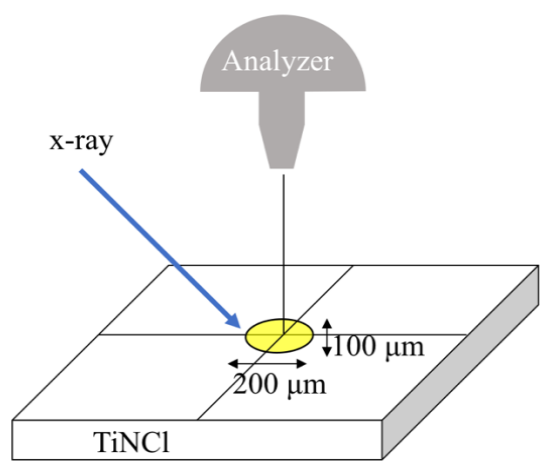

Figure 1. Schematic diagram of the x-ray irradiation.

**Results and discussion**

Figure 2 shows the soft x-ray irradiation time dependence of the valence band PES

spectra of TiNCl. The background of an iterative Shirley method [41, 42] was subtracted from the raw spectra. The initial valence band spectrum of TiNCl had a peak at 6 eV with a shoulder structure at 4 eV, and there was almost no intensity in the near-$E_F$ region. The overall spectral shape is consistent with that of the previous study [25], and thus the states of approximately 6 and 4 eV can be ascribed to Cl $3p$ and N $2p$ orbitals, respectively.

The peak intensities at 6 eV and 4 eV were decreased gradually. In contrast, the spectral intensity in the region between $E_F$ and ~ 1.5 eV appears to be increased, as illustrated in the inset that shows a blowup of the spectra near $E_F$. The intensity near $E_F$ of the initial spectrum can be ascribed to the states induced by the soft x-ray irradiation during the measurement and/or may be assigned to impurity states. In the spectrum after 150 min irradiation, a clear Fermi edge structure was observed, suggesting the metallic nature of the measured region of the sample surface. The change in the valence band and the intensity near $E_F$ due to the soft x-ray irradiation is very similar to the change due to the photoirradiation effect of $SrTiO_3$ [30].

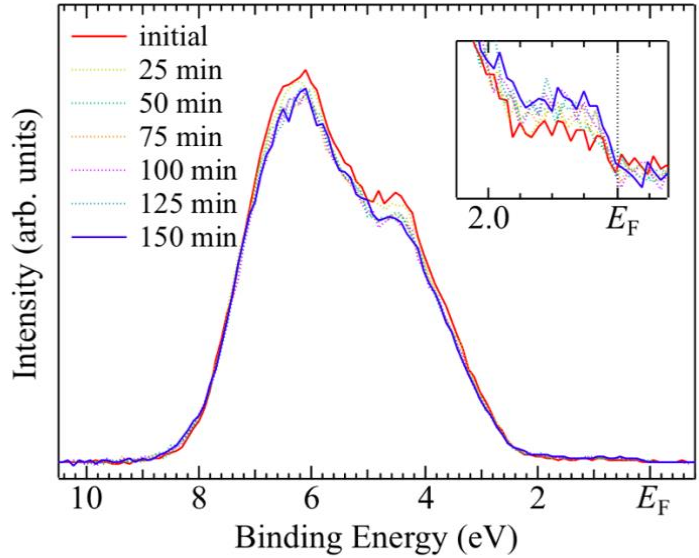

Figure 2. Soft x-ray irradiation time dependence of the valence band PES spectra of TiNCl, measured with a photon energy of 1000 eV at 100 K. Inset shows the enlargement of the spectra near $E_F$.

To spectroscopically confirm the metallization and the character of the states at $E_F$, we performed RPES of irradiated TiNCl at the Ti $2p$-$3d$ absorption threshold. The photon energies for the RPES measurement were chosen based on the Ti $L_3$ XAS spectrum, as shown in the inset of Fig. 3. Ti $L_3$ XAS spectrum had sharp absorption peaks at 457.4 eV, 458.5 eV, and 460.1 eV in the region of Ti $2p_{3/2}$-$3d$ absorption edge. In the octahedral structure, the 3$d$ level was split into $e_g$ and $t_{2g}$ states by the crystal field effect. However, since the Ti ions in TiNCl are in a distorted octahedral structure [7], the degeneracy was further removed by the static Jahn-Teller effect, leading to the observed three-peak structure. We measured RPES spectra of TiNCl in the off-resonance ($h\nu$ = 454.0 eV) and on-resonance ($h\nu$ = 457.4 eV) conditions after the irradiation the soft x-rays for 4 h. While the spectral intensity at $E_F$ in the off-resonance spectrum was almost negligible, the one in the on-resonance spectrum exhibited a distinctive Fermi edge, and was considerably enhanced. The resonance enhancement of the near-$E_F$ structure indicates that its orbital character is Ti 3$d$. Furthermore, the presence of the distinctive Fermi edge indicates that the soft x-ray irradiation indeed induces the metallization of TiNCl.

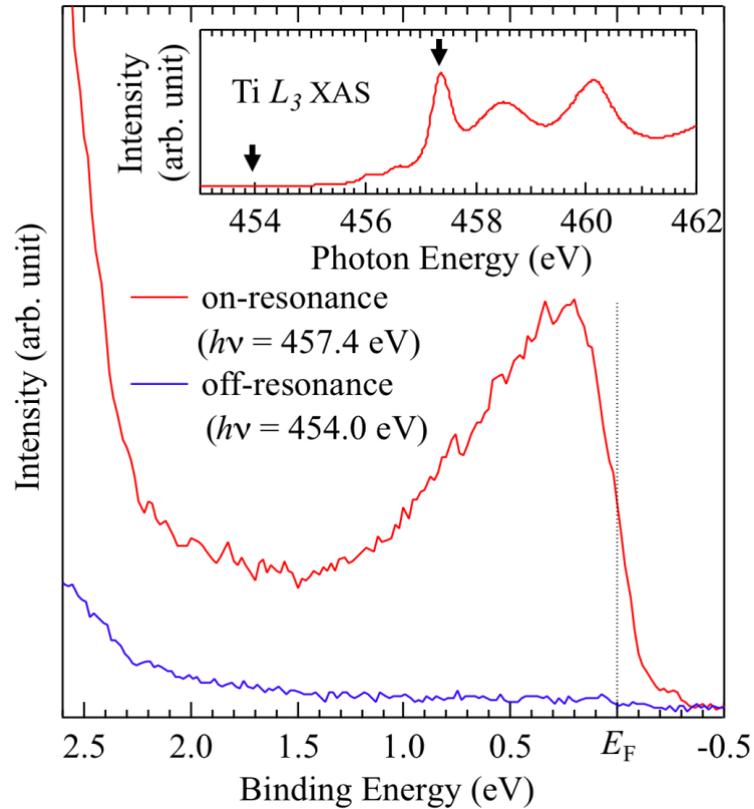

Figure 3. ON (red) and OFF (blue) Ti 2$p$-3$d$ resonance photoemission spectra of the irradiated TiNCl measured with a photon energy of 457.4 eV and 454.0 eV, respectively. The inset shows the Ti 2$p$ absorption spectrum where the employed photon energies for the resonance photoemission are indicated by arrows.

To understand the mechanism of the metallization induced by soft x-ray irradiation in terms of chemical states, we have measured the core level spectra of TiNCl with different irradiation conditions. Figure 4 (a) shows soft x-ray irradiation time dependence of the Ti 2$p$ core-level spectra, and Fig.4 (b) represents the initial and 150 min difference spectrum.

The peaks at 457.5 eV and 463.5 eV are spin-orbit split Ti 2*p*, namely Ti 2$p_{3/2}$ and Ti 2$p_{1/2}$, respectively [43]. The initial Ti 2*p* spectrum showed a sharp Ti$_{4+}$ peak at 457.5 eV [25] and a small shoulder structure at 456.5 eV. In the Ti 2*p* spectrum, as the irradiation time increased, the peak area of Ti$_{4+}$ decreased compared to the initial value. In addition, the intensity of the shoulder structure at 456.5 eV clearly increased gradually. This binding energy is different from that of TiN (455.2 ± 0.2 eV) [44, 45], but is very similar to the week Ti 2*p* spectral component of electron-doped SrTiO$_3$ and electron-doped anatase-TiO$_2$, which is assigned to the component of Ti$_{3+}$ state [30, 31, 46-48]. Thus, the observation indicates that the soft x-ray irradiation on TiNCl induces the changes Ti$_{4+}$ in the valence state of the Ti ion from the Ti$_{3+}$ state to the Ti$_{4+}$ state. We performed spectral fitting using two components. The area intensities of Ti$_{3+}$ and Ti$_{4+}$ components at each irradiation time are summarized in Table 1.

Figure 4 (c) compares the spectra of the N 1*s* core level at each soft x-ray irradiation time. These spectra had peaks at 396.5 eV, and were almost identical, suggesting that there was no significant change in the spectral shape due to soft x-ray irradiation. The Cl 2*p* core-level spectra consist of Cl 2$p_{3/2}$ (198.4 eV) and Cl 2$p_{1/2}$ (200.1 eV) spin-orbit split peaks. The intensity of the Cl 2*p* spectrum after the soft x-ray irradiation of 150 min decreased gradually, which suggests the desorption of Cl atoms from the surface is induced by the soft x-ray irradiation. This is consistent with the observed reduction of the valence band spectra, which reflects a dominant Cl 3*p* contribution.

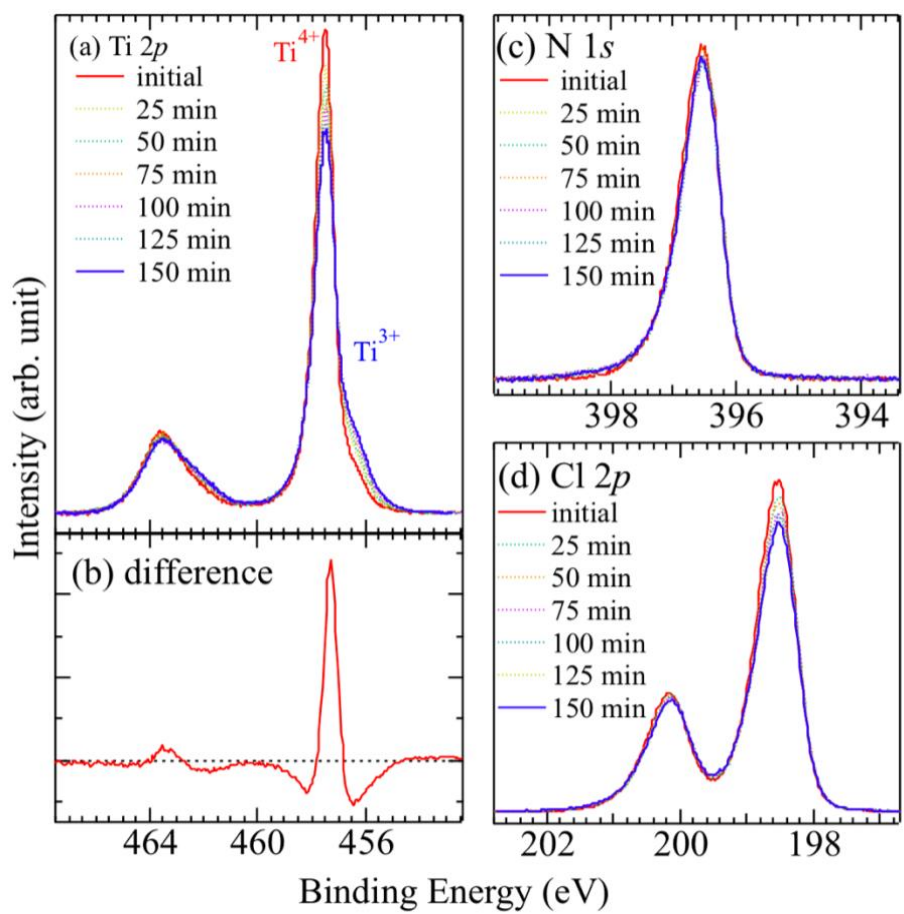

Figure 4. Soft x-ray irradiation time dependence of core-levels PES spectra. (a) Ti 2p core-level spectra irradiated from the initial to 150 min in the interval of 25 min. (b) Difference between the initial and 150min. (c) and (d) are analogous spectra to (a) for the N 1s and Cl 2p core-level, respectively.

Table 1. Area intensity ratio estimated from $Ti_{3+}$ and $Ti_{4+}$ in Ti 2p core level PES spectra at each irradiation time.

| Irradiation time (min) | Relative area intensity of components (%) | |
|---|---|---|
| | $Ti^{4+}$ in Ti 2p | $Ti^{3+}$ in Ti 2p |
| 0 | 90.60 | 9.40 |
| 25 | 87.37 | 12.63 |
| 50 | 80.22 | 19.78 |
| 75 | 77.52 | 22.48 |
| 100 | 74.26 | 25.74 |
| 125 | 73.21 | 26.79 |
| 150 | 70.96 | 29.04 |

Here, we discuss the relationship between the observed results. Figure 5 shows the soft x-ray irradiation time dependence of the intensity of N 1$s$ and Cl 2$p$ core-level PES spectra, the PES intensity between $E_F$ and 1.5 eV, and the Ti$_{3+}$ ratio (Ti$_{3+}$ / ( Ti$_{4+}$ + Ti$_{3+}$ )) of Ti 2$p$ core-level PES spectrum. We normalized their initial values to unity. The intensity of N 1$s$ core-level PES does not show a marked change depending on the soft x-ray irradiation time of soft x-ray. In a comparison with that of the N 1$s$ core-level PES, the intensity of the Cl 2$p$ core-level PES decreased more rapidly with increasing irradiation time. Furthermore, simultaneously with a decrease in Cl, the areas of the Ti$_{3+}$ component and of the states near $E_F$ increased in a similar manner. A smaller increase in near $E_F$ intensity may be attributed to a more complicated orbital character of states near $E_F$ than that of the Ti$_{3+}$ core level. These observations suggest the close relationship of the desorption of Cl atoms and an increase in the intensity near $E_F$ with an increase in the Ti$_{3+}$ component. We consider that the desorption of a portion of the Cl atoms from the TiNCl layers introduces electron carriers into the layers, making the layer conductive, which is inferred from the observation of the metallic Fermi edge. Therefore, to estimate the time of change, the data for Ti$_{3+}$ and Cl 2$p$, which are more intense than the near-$E_F$ data, were fitted with an exponential function ($A\exp(-t/\lambda)+B$ for Ti$_{3+}$ and $C\exp(t/\lambda)+D$ for Cl 2$p$, where $t$ and $\lambda$ are time and lifetime, respectively.). The obtained lifetimes of Ti$_{3+}$ and Cl 2$p$ were 119 $\pm$ 29 and 96 $\pm$ 31 min, respectively. This provided a value of 100 min as the order estimate of the time of change for this experiment.

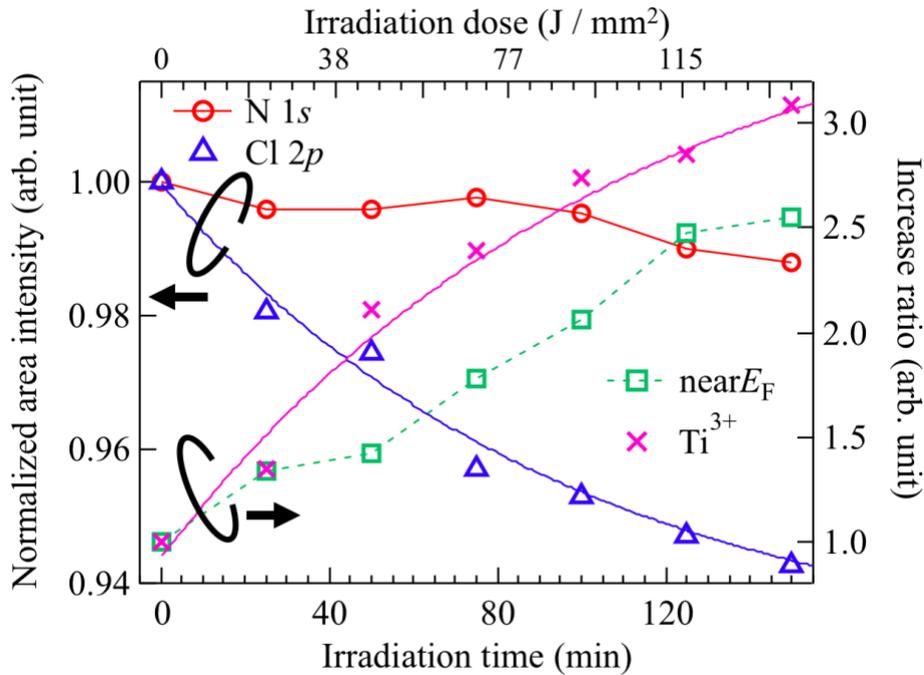

Figure 5. Soft x-ray irradiation time dependence of the core-level and near-$E_F$ intensities of TiNCl. Area intensity of N 1$s$ and Cl 2$p$ core-level spectra (left axis), and an increase ratio in area intensity of near $E_F$ and Ti$_{3+}$ in Ti 2$p$ core level spectra (right axis). The initial values were normalized to 1.

There are two possible causes of the desorption of Cl atoms. One of them is a pyrolysis. TiNCl has been reported to be decomposed completely into TiN when it is heated above 550°C [27-29]. In the present experiment, however, the sample was kept at 100 K during the measurement, and therefore this possibility can be ruled out. The other possible cause is photon stimulated desorption [49-51]. There are primarily two mechanisms for photon stimulated desorption, which are distinguished by the process induced by photon irradiation: The Menzel, Gomer, and Redhead (MGR) model and the Knotek and Feibelman (KF) model [49, 50]. The MGR model involves the excitation of valence electrons and works reasonably well for covalent bonds. The KF model involves the excitation of core electrons and is applicable to ion bonding. Since the TiNCl is considered to be an ionic insulator composed of TiN$_+$ and Cl$_-$ [23], we assumed that this mechanism is involved in the Cl desorption of TiNCl. In the case of TiO$_2$, photoinduced desorption is explained by the KF model in terms of Coulomb expulsion between Ti$_{4+}$ and O$_+$ that are induced by emitting Auger electrons after the relaxation of electrons in O 2$p$ levels to create photo-holes in the Ti 3$p$ levels [48]. If the KF mechanism is involved in the present case, the electron transfer that leads to Coulomb expulsion may occur from Cl 3$p$ (valence band) to Ti 3$p$. Further experiments are needed to confirm the KF mechanism, as it is we think beyond the scope of the present study.

With photon stimulated Cl desorption, the final products may be TiN or Cl deficient TiNCl$_{1-\delta}$, depending on the amount of Cl deficiency. However, from the Ti 2$p$ photoelectron spectrum, no sharp TiN peak was observed at 455±0.3 eV[44, 45], suggesting the formation of TiNCl$_{1-\delta}$, rather than TiN. In β-(Hf or Zr) NCl, Cl deficiency is called deintercalation, which is one of the methods of electron doping to the system. Deintercalated HfNCl$_{0.7}$ is a metallic system that also exhibits superconductivity below 24 K [26]. However, since electron doping utilizing the Cl deintercalation into α-$M$N$X$ has never been reported, the present study is the first experimental realization of deintercalated electron doped α-$M$N$X$. Further Cl deintercalation by the soft x-ray irradiation would make the system superconductive, like β-form compounds.

Lastly, we discuss the benefits of methodologies and findings. Regarding pyrolysis, bond breaking by heat cannot be confined within a small area because of the diffusion of heat in a solid. In addition, temperature controlling of a sample is severe for a target material located in a narrow temperature region of a complicated phase diagram. In contrast, bond breaking by light irradiation occurs only in the place where light is irradiated. In particular, when core-levels excitation are important in the bond-breaking process as in the KF model, element/chemical site-selective bond breaking is possible by setting the excitation energy to a specific element/chemical site [52, 53]. This site-selective bond breaking is called a molecular scalpel and is mainly studied in surface adsorption molecular systems. Such a characteristic, combined with sub-micrometer scale spot sizes in third-generation synchrotron facilities, might be used to make sub-micron scale conducting paths at any place on the surface. Photo-induced metallization of TiNCl has enabled detailed electronic-structure investigation studies of metallic (even superconductive) TiNCl by controlling the carrier concentration in a small single crystal region on the surface, which shed light on the exotic properties of TiNCl.

**Conclusion**

In summary, the irradiation time dependence of the valence band and core-level spectra of TiNCl was studied by PES and Ti $2p$-$3d$ RPES. Using soft x-ray irradiation, the intensity of the valence band structure origination from Cl $3p$ and N $2p$ states decreased while the intensity near $E_F$ increased. The intensity near $E_F$ exhibited resonant enhancement, and the Fermi edge was clearly observed by Ti $2p$-$3d$ RPES. The analyses of core-level and near-$E_F$ PES spectral intensities as a function of the soft x-ray irradiation time revealed that the desorption of Cl atoms, the increase in the intensity of the Ti$_{3+}$ component, and increase of the spectral weight in the vicinity of $E_F$ exhibit strong correlations. These results indicate that photoirradiation induced metallization of TiNCl occurs due to electron doping through Cl desorption. As estimated from Cl $2p$ and Ti$_{3+}$ data, the order of the time-of-changes is 100 min. This technique can be used to further explore the conducting properties, especially the unconventional superconductivity, of TiNCl.

**Acknowledgment**

This work was supported by the Fund for the Promotion of Joint International Research (B) (No. 18KK0076) from the Ministry of Education, Culture, Sports, Science and Technology of Japan (MEXT), and JSPS KAKENHI Grant Number JP18K04707. This work was performed under the Shared Use Program of Japan Atomic Energy Agency (JAEA) Facilities (Proposal No. 2019A-E19) with the approval of Nanotechnology Platform project supported by the Ministry of Education, Culture, Sports, Science and Technology (Proposal No. A-19-AE-0019). The synchrotron radiation experiments were performed at JAEA beamline BL23SU in SPring-8 (Proposal No.2019A3844). The authors would like to thank Emeritus Professor Shoji Yamanaka of Hiroshima University for his helpful suggestion in the synthesis of samples.

References

1. Yamanaka S. 2000 HIGH-$T_c$ SUPERCONDUCTIVITY IN ELECTRON-DOPED LAYER STRUCTURED NITRIDES *Annu. Rev. Mater. Sci.* **30** 53

2. Juza R, Heners J 1964 Über Nitridhalogenide des Titans und Zirkons *Z. Anorg. Allg. Chem.* **332** 159

3. Woodward PM, Vogt T 1998 Electronic Band Structure Calculations of the *MNX* (*M*=Zr, Ti; *X*=Cl, Br, I) System and Its Superconducting Member, Li-Doped *β*-ZrNCl *J. Solid State Chem.* **138** 207

4. Liang Y, Dai Y, Ma Y, Ju L, Wei W, Huang B 2018 Novel titanium nitride halide TiNX (X=F, Cl, Br) monolayers: potential materials for highly efficient excitonic solar cells *J. Mater. Chem.* A **6** 2073

5. Zhou L, Zhuo Z, Kou L, Du A, Tretiak S 2017 Computational Dissection of Two-Dimensional Rectangular Titanium Mononitride TiN: Auxetics and Promises for Photocatalysis *Nano Lett.* **17** 4466

6. Wang A, Wang Z, Du A and Zhao M 2016 Band inversion and topological aspects in a TiNI monolayer *Phys. Chem. Chem. Phys.* **18** 22154

7. Yamanaka S, Yasunaga T, Yamaguchi K, Tagawa M 2009 Structure and superconductivity of the intercalation compounds of TiNCl with pyridine and alkali metals as intercalants *J. Mater. Chem.* **19** 2573

8. Yamanaka S, Kawaji H, Hotehama K, Ohashi M 1996 A New Layer-Structured Nitride Superconductor. Lithium-Intercalated β-Zirconium Nitride Chloride, Li$_x$ZrNCl *Adv. Mater.* **8** 771

9. Yamanaka S, Hotehama K, Kawaji H. Nature. 1998 Superconductivity at 25.5 K in electron-doped layered hafnium nitride **392** 580

10. Tou H, Maniwa Y, Koiwasaki T, Yamanaka S 2001 Unconventional Superconductivity in Electron-Doped Layered Li$_{0.48}$(THF)$_y$HfNCl *Phys. Rev. Lett.* **86** 5775

11. Yokoya T, Ishiwata Y, Shin S, Shamoto S, Iizawa K, Kajitani T *et al* 2001 Changes of electronic structure across the insulator-to-metal transition of quasi-two-dimensional Na-intercalated *β*-HfNCl studied by photoemission and x-ray absorption *Phys. Rev.* B **64** 153107

12. Takeuchi T, Tsuda S, Yokoya T, Tsukamoto T, Shin S, Hirai A *et al* 2003 Soft X-ray emission and high-resolution photoemission study of quasi-two-dimensional superconductor Na$_x$HfNCl *Physica* C **392–396** 127

13. Yokoya T, Takeuchi T, Tsuda S, Kiss T, Higuchi T, Shin S *et al* 2004 Valence-band photoemission study of *β*-ZrNCl and the quasi-two-dimensional superconductor

Na$_x$ZrNCl *Phys. Rev.* B. **70** 193103

14. Tou H, Maniwa Y, Yamanaka S 2003 Superconducting characteristics in electron-doped layered hafnium nitride: 15N isotope effect studies *Phys. Rev.* B **67** 100509(R)
15. Taguchi Y, Kitora A, Iwasa Y 2006 Increase in $T_c$ upon Reduction of Doping in Li$_x$ZrNCl Superconductors *Phys. Rev. Lett.* **97** 107001
16. Takano T, Kishiume T, Taguchi Y, Isawa Y 2008 Interlayer-Spacing Dependence of $T_c$ in Li$_x$M$_y$HfNCl (*M*: Molecule) Superconductors *Phys. Rev. Lett.* **100** 247005
17. Kasahara Y, Kishiume T, Kobayashi K, Taguchi Y, Iwasa Y 2010 Superconductivity in molecule-intercalated Li$_x$ZrNCl with variable interlayer spacing *Phys. Rev.* B **82** 054504
18. Kuroki K 2010 Spin-fluctuation-mediated *d+id'* pairing mechanism in doped *β-M*NCl (*M*=Hf, Zr) superconductors *Phys. Rev.* B **81** 104502
19. Saito Y, Kasahara Y, Ye J, Iwasa Y, Nojima T 2015 Metallic ground state in an ion-gated two-dimensional superconductor Science **350** 409
20. Nakagawa Y, Saito Y, Nojima T, Inumaru K, Yamanaka S, Kasahara Y *et al* 2018 Gated-controlled low carrier density superconductors: Toward the two-dimensional BCS-BEC crossover *Phys. Rev.* B **98** 064512
21. Zhang S, Tanaka M, Yamanaka S 2012 Superconductivity in electron-doped layered TiNCl with variable interlayer coupling *Phys. Rev.* B **86** 024516
22. Sugimoto A, Shohara K, Ekino T, Zheng Z, and Yamanaka S 2012 Nanoscale electronic structure of the layered nitride superconductors *α*-K$_x$TiNCl and *β*-HfNCl$_y$ observed by scanning tunneling microscopy and spectroscopy *Phys. Rev.* B **85** 144517
23. Yin Q, Ylvisaker ER, Pickett WE 2011 Spin and charge fluctuations in *α*-structure layered nitride superconductors *Phys. Rev.* B **83** 014509
24. Harshman DR, Fiory AT 2014 Comment on "Superconductivity in electron-doped layered TiNCl with variable interlayer coupling" *Phys. Rev.* B **90** 186501
25. Kataoka N, Terashima K, Tanaka M, Hosoda W, Taniguchi T, Wakita T *et al* 2019 µ-PES Studies on TiNCl and Quasi-two-dimensional Superconductor Na-intercalated TiNCl *J. Phys. Soc. Jpn.* **88** 104709
26. Zhu L, Yamanaka S 2003 Preparation and Superconductivity of Chlorine-Deintercalated Crystals *β*-MNCl$_{1-x}$ (M=Zr, Hf) *Chem. Mater.* **15** 1897
27. Saeki Y, Matsuzaki R, Yajima A, Akiyama M 1982 Reaction Process of Titanium Tetrachloride with Ammonia in the Vaper Phase and Properties of the Titanium Nitride Formed *Bull. Chem. Soc. Jpn.* **55** 3193
28. Sosnov E, Malkov A, Malygin A. Russ 2015 Chemical Transformations at the Silica

Surface upon Sequential Interactions with Titanium Tetrachloride and Ammonia Vapors *J. Gen. Chem.* **85** 2533

29. Hegde RI, Fiordalice RW, Tobin PJ 1993 TiNCl formation during low-temperature, low-pressure chemical vapor deposition of TiN *Appl. Phys. Lett.* **62** 2326
30. Plumb NC *et al* 2014 Mixed Dimensionality of Confined Conducting Electrons in the Surface Region of $SrTiO_3$ *Phys. Rev. Lett.* **113** 086801
31. Plumb NC *et al* 2017 Evolution of the $SrTiO_3$ surface electronic state as a function of $LaAlO_3$ overlayer thickness *Appl. Surf. Sci.* **412** 271
32. Meevasana W *et al* 2011 Creation and control of a two-dimensional electron liquid at the bare $SrTiO_3$ surface *Nat. Mater.* **10** 114
33. Reckers P *et al* 2012 Deep and Shallow $TiO_2$ Gap States on Cleaved Anatase Single Crystal (101) Surfaces, Nanocrystalline Anatase Films, and ALD Titania Ante and Post Annealing *J. Phys. Chem.* C. **119** 9890
34. Rödel TC *et al* 2015 Engineering two-dimensional electron gases at the (001) and (101) surfaces of $TiO_2$ anatase using light *Phys. Rev.* B **92** 041106(R)
35. Wang Z *et al* 2017 Atomically Precise Lateral Modulation of a Two-Dimensional Electron Liquid in Anatase $TiO_2$ Thin Films *Nano Lett.* **17** 2561
36. King P *et al* 2012 Subband Structure of a Two-Dimensional Electron Gas Formed at the Polar Surface of the Strong Spin-Orbit Perovskite $KTiO_3$ *Phys. Rev. Lett.* **108** 117602
37. Santander-Syro A *et al* 2011 Two-dimensional electron gas with universal subbands at the surface of $SrTiO_3$ **469** 189
38. Santander-Syro A *et al* 2014 Giant spin splitting of the two-dimensional electron gas at the surface of $SrTiO_3$ *Nature Nat. Mater.* **13** 1085
39. King P *et al* 2014 Quasiparticle dynamics and spin–orbital texture of the $SrTiO_3$ two-dimensional electron gas *Nat. Commun.* **5** 3414
40. Saitoh Y *et al* 2012 Performance upgrade in the JAEA actinide science beamline BL23SU at SPring-8 with a new twin-helical undulator *J. Synchrotron Rad.* **19** 388
41. Shirley D 1972 High-Resolution X-Ray Photoemission Spectrum of the Valence Bands of Gold *Phys. Rev.* B **5** 4709
42. Proctor A, Sherwood P 1982 Data Analysis Techniques in X-ray Photoelectron Spectroscopy *Anal. Chem.* **54** 13
43. The binding energies of Ti $2p$ and Cl $2p$ of the present study are slightly lower by 0.4 eV and 0.2 eV than those in Ref. 25, respectively. The two experiments were performed in SPring-8 but on the different beam lines BL23SU (present study) and BL25SU (Ref. 25). We assume that there are three reasons for this difference:

charging up, difference in metallization, and calibration of the spectrometer. Since the photon flux density of BL25SU is higher than that of BL23SU, larger binding energies of BL25SU data may be explained by charging up and/or difference in metallization. We do not have the clear answer, however the difference does not change the conclusion of the present study.

44. Saha NC, Tompkins HG 1992 Titanium nitride oxidation chemistry: An x-ray photoelectron spectroscopy study *J. Appl. Phys.* **72** 3072
45. Jaeger D, Patscheider J 2012 A complete and self-consistent evaluation of XPS spectra of TiN *J. Electron Spectrosc. Relat. Phenom.* **185** 523
46. Ishida Y *et al* 2008 Coherent and Incoherent Excitations of Electron-Doped $SrTiO_3$ *Phys. Rev. Lett.* **100** 056401
47. Sing M *et al* 2009 Profiling the Interface Electron Gas of $LaAlO_3/SrTiO_3$ Heterostructures with Hard X-Ray Photoelectron Spectroscopy *Phys. Rev. Lett.* **102** 176805
48. Yukawa R *et al* 2018 Control of two-dimensional electronic states at anatase $TiO_2$ (001) surface by K adsorption *Phys. Rev. B* **97** 165428
49. Knotek ML, Feibelman PJ 1978 Ion Desorption by Core-Hole AugerDecay *Phys. Rev. Lett.* **40** 964
50. Ramaker DE, White CT, Murday JS 1982 ON AUGER INDUCED DECOMPOSITION/DESORPTION OF COVALENT AND IONIC SYSTEMS *Phys. Lett.* A **89** 211
51. Segovia J 1995 A review of electron stimulated desorption processes influencing the measurement of pressure or gas composition in ultra high vacuum systems *Vacuum* **47** 333
52. Eberhardt W *et al* 1983 Site-Specific Fragmentation of Small Molecules Following Soft-X-Ray Excitation *Phys. Rev. Lett.* **50** 1957
53. Wada S *et al* 2006 Selective chemical bond breaking characteristically induced by resonant core excitation of ester compounds on a surface *J. Phys.: Condens. Matter* **18** S 1629